# Chemical Bonding in Epitaxial ZrB$_2$ Studied by X-ray Spectroscopy

Martin Magnuson, Lina Tengdelius, Grzegorz Greczynski, Lars Hultman, and Hans Högberg

*Thin Film Physics Division, Department of Physics, Chemistry and Biology (IFM),*
*Linköping University, SE-58183 Linköping, Sweden*

## Abstract

The chemical bonding in an epitaxial ZrB$_2$ film is investigated by Zr *K*-edge (1s) X-ray absorption near-edge structure (XANES) and extended X-ray absorption fine structure (EXAFS) spectroscopies and compared to the ZrB$_2$ compound target from which the film was synthesized as well as a bulk α-Zr reference. Quantitative analysis of X-ray Photoelectron Spectroscopy spectra reveals at the surface: ~5% O in the epitaxial ZrB$_2$ film, ~19% O in the ZrB$_2$ compound target and ~22% O in the bulk α-Zr reference after completed sputter cleaning. For the ZrB$_2$ compound target, X-ray diffraction (XRD) shows weak but visible $\bar{1}11$, 111, and 220 peaks from monoclinic ZrO$_2$ together with peaks from ZrB$_2$ and where the intensity distribution for the ZrB$_2$ peaks show a randomly oriented target material. For the bulk α-Zr reference no peaks from any crystalline oxide were visible in the diffractogram recorded from the 0001-oriented metal. The Zr *K*-edge absorption from the two ZrB$_2$ samples demonstrate more pronounced oscillations for the epitaxial ZrB$_2$ film than in the bulk ZrB$_2$ attributed to the high atomic ordering within the columns of the film. The XANES exhibits no pre-peak due to lack of *p-d* hybridization in ZrB$_2$, but with a chemical shift towards higher energy of 4 eV in the film and 6 eV for the bulk compared to α-Zr (17.993 keV) from the charge-transfer from Zr to B. The 2 eV larger shift in bulk ZrB$_2$ material suggests higher oxygen content than in the epitaxial film, which is supported by XPS. In EXAFS, the modelled cell-edge in ZrB$_2$ is slightly smaller in the thin film (*a*=3.165 Å, *c*=3.520 Å) in comparison to the bulk target material (*a*=3.175 Å, *c*=3.540 Å) while in hexagonal closest-packed metal (α-phase, *a*=3.254 Å, *c*=5.147 Å). The modelled coordination numbers show that the EXAFS spectra of the epitaxial ZrB$_2$ film is highly anisotropic with strong in-plane contribution, while the bulk target material is more isotropic. The Zr-B distance in the film of 2.539 Å is in agreement with the calculated value from XRD data of 2.542 Å. This is slightly shorter compared to that in the ZrB$_2$ compound target 2.599 Å, supporting the XANES results of a higher atomic order within the columns of the film compared to bulk ZrB$_2$.

## 1. Introduction

Transition metal borides with an AlB$_2$ type structure (Strukturbericht notation C32) are an emerging class of thin film materials that are much less investigated compared to hard and refractory carbides and nitrides. Among these borides, ZrB$_2$, demonstrates properties favorable for metal cutting, aerospace or electronic applications. This is due to the materials property envelope including high melting point (3245 °C), high hardness (23 GPa), and good corrosion resistance in combination with the highest electrical conductivity of borides with a C32 structure seen from a value of $1.0 \times 10^7$ Sm$^{-1}$ [1]. Hence, the crystal structure is the key to the properties exhibited by ZrB$_2$. The unit cell of ZrB$_2$ with Zr atoms at the basis (0,0,0) and boron atoms positioned in the trigonal prismatic interstitials at (⅓, ⅔, ½) and (⅔, ⅓, ½). From translating the unit cell of ZrB$_2$, the symmetry in the boride becomes evident, where the B atoms form honeycombed, graphite-like sheets that are interleaved between hexagonal close packed Zr layers. Charge-transfer occurs from Zr to B to stabilize the B-B sheet, which results in an anisotropic electronic structure and chemical bonding. This is different from the hard and refractory transition metal carbides and nitrides with their NaCl type structure (Strukturbericht notation B1) that are isotropic materials and in contrast to borides, the C or N atoms are separated in the crystal structure.

Unlike growth of hard and refractory carbide and nitride films there is no reactive sputtering process for borides such as ZrB$_2$. Consequently, sputtering of ZrB$_2$ films are conducted from ZrB$_2$ compound targets or from Zr-B composite targets. Recently, we advanced sputtering of ZrB$_2$ films by demonstrating epitaxial growth on Si(111) [2], 4H-SiC(0001) [2,3], and Al$_2$O$_3$(0001) substrates [4], using a substrate temperature of 900 °C. On investigated substrate materials, the ZrB$_2$ films grow as epitaxial columns a few to some tens of nm wide parallel to the substrate and extend all the way to the film-vacuum interface orthogonal to the substrate. The deposition of ZrB$_2$ films is typically troubled by growth of amorphous films and fine-grained films with composition that deviates from stoichiometry as well as growth of films with a high level of contaminants than foremost oxygen, see e.g. [5-12]. In order to advance the synthesis and control properties of this interesting boride, the chemical bonding needs deeper understanding. Comparison between epitaxial films and bulk polycrystalline samples has been lacking in literature.

X-ray absorption near-edge spectroscopy (XANES) and extended X-ray absorption fine structure (EXAFS) are ideal techniques for determining the electronic structure properties of ZrB$_2$ including thin films. Both techniques have the advantage of being element specific, with signals like fingerprints, in terms of coordination chemistry around the Zr atoms. For ZrB$_2$, mainly bulk materials have been studied and with one study on thin films. Chu *et al.* [13] reported on EXAFS on polycrystalline ZrB$_2$ samples synthesized from Zr and B powders by floating zone method. The temperature-dependence showed little difference between the in-plane and out-of-plane vibrations of the Zr-Zr bonds [14]. On the other hand, Bösenberg *et al.* [15] investigated the chemical states of ZrB$_2$ and Zr powders by both XANES and EXAFS. The Zr edge energy of the ZrB$_2$ was found to be 10 eV above that of the Zr foil reference. For e-beam co-evaporated ZrB$_2$ thin films, EXAFS and XANES have been applied to investigate the atomic distances and the local chemical bonding structure [16]. The oxygen content in these films resulted in the formation of crystalline tetragonal ZrO$_2$, which yielded longer atomic bond distances as determined by EXAFS while XANES showed high-energy shifted unoccupied Zr *4d* electronic states.

In this work, we investigate the properties of a ~1100 nm thick epitaxial ZrB$_2$ film deposited on Al$_2$O$_3$(0001) substrate by magnetron sputtering (DCMS), using a combination of Zr *K*-edge (*1s*) XANES and EXAFS spectroscopies. The local chemical bonding structure and atomic distances in the epitaxial film is compared to these properties determined for the $ZrB_2$ compound target from which the film was deposited from as well as to a bulk α-Zr reference. The XANES and EXAFS measurements are supported by X-ray Photoelectron Spectroscopy (XPS) for composition and chemical bonding structure and X-ray diffraction (XRD) for structural properties and atomic distances.

## 2. Experimental Details

**2.1 Thin film deposition and fundamental characterization**
The investigated epitaxial $ZrB_2$ film was deposited on a $Al_2O_3(0001)$ substrate by DCMS from a $ZrB_2$ compound target 99.5% purity from Kurt J. Lesker Company, Clairton, PA, USA using a laboratory scale ultrahigh-vacuum deposition system. The film was grown to a thickness of ~1100 nm at a substrate temperature of 900 °C with a sputtering power of 400W on the three-inch circular sputtering source and in an argon plasma (99.9997%) held at a 0.53 Pa. For further details of the process conditions, the reader is referred to ref. [2] and for the properties of the films to ref. [4]. The investigated α-Zr bulk reference was a commercial zirconium target with a purity of 99.9% from Kurt J. Lesker Company, Clairton, PA, USA. The thickness of the target materials was 3175 μm (1/8 inch).

Analysis by XPS was conducted in a Kratos AXIS UltraDLD, Manchester, U.K. system with monochromatic Al-$K_α$ (1486.6 eV) radiation to determine the composition and chemical bonding structure of the epitaxial $ZrB_2$ film, the $ZrB_2$ compound target, and the bulk α -Zr reference. The samples were analyzed both in the as-deposited state and after sputter-cleaning with 4 keV $Ar^+$ ions incident at an angle of 70° with respect to the surface normal, to remove adsorbed contaminants following air exposure. The sputtered area was 3x3 $mm^2$ and the analyzed area was 0.3x0.7 $mm^2$. The epitaxial film was analyzed after sputter-cleaning for 60, 120, and 180 s, while the target and the reference were analyzed after 120, 240, 360, 480, 600, and 720 s sputter-cleaning. The binding energy scale was calibrated against the Fermi level cut-off using the procedure described in detail elsewhere [17]. In this way, the uncertainties associated with using the C *1s* peak of adventitious carbon for calibration of the binding energy scale [18] are avoided. Quantification of the elements in the samples was performed using Casa XPS software (version 2.3.16), based upon peak areas from narrow energy range scans and elemental sensitivity factors supplied by Kratos Analytical Ltd. [19]. The quantification accuracy of XPS is typically around ± 5 %. The structural properties of the films were assessed by X-ray diffraction (XRD) performing θ/2θ scans in a Philips PW 1820 Bragg-Brentano diffractometer using Cu $K_α$ radiation at settings of 40 kV and 40 mA. XRD pole figures were recorded in a PANalytical EMPYREAN diffractometer at 45 kV and 40 mA to determine the epitaxial growth relationships of the $ZrB_2$ film to the $Al_2O_3(0001)$ substrate.

**2.2 XANES and EXAFS measurements**
The XANES and EXAFS spectra were measured at the undulator beamline I811 on the MAX II ring of the MAX IV Laboratory, Lund University, Sweden [20]. The energy resolution at the Zr *1s* edge of the beamline monochromator was 0.5 eV. The X-ray absorption spectra were recorded in reflection mode by detecting the fluorescence yield [21], using a passivated implanted planar silicon (PIPS) detector from 150 eV below to 1200 eV above the Zr *1s* absorption edge energy ($E_0$=17.993 keV) with 0.5 eV energy steps. To minimize self-absorption effects in the sample and Bragg scattering from the substrate, the incidence angle on the sample was normal to the sample surface and varied in 0.25° steps in a maximum range of ±3° using a stepper motor. For the thick and concentrated bulk reference samples, the extracted values of the coordination numbers represent a lower limit due to self-absorption effects, while their bond lengths are not affected. Out-of-plane measurements were not included in this study as in the grazing incidence geometry, there major self-absorption effects can be observed resulting in reduced amplitudes and significantly lower signal-to-noise ratio.

Based on the fitting results, the Zr-B, and Zr-Zr scattering paths obtained from the Effective Scattering Amplitudes (FEFF) [22,23] were included in the EXAFS fitting procedure using the *Visual Processing in EXAFS Researches* (VIPER) software package [24]. The threshold energy $E_o$, is defined through the point of inflection of the absorption edge. In the fitting procedure, $E_o$ is used as an adjustable parameter that partly compensate for errors in the phase shifts. The edge reference energy $E_0$ was set to the pre-peak of each Zr *1s* X-ray absorption spectrum as determined from the first peak of the derivative of each spectrum relative to the pure hexagonal Zr reference sample $E_0$ was set to 17.99273 keV (α-Zr, $x=0$).

The $k^2$-weighted χ EXAFS oscillations were extracted from the raw absorption data after removing known monochromator-induced glitches and peaks originating from substrate diffraction, subsequent atomic background subtraction, and averaging of 15 absorption spectra. The bond distances ($R$), number of neighbors ($N$), Debye-Waller factors ($\sigma^2$, representing the amount of disorder) and the reduced $\chi_r^2$ as the squared area of the residual, were determined by fitting the back-Fourier-transform signal between $k$=3-12 Å$^{-1}$ originally obtained from the forward Fourier-transform within $R$=2-3.5 Å of the first coordination shell using a Hanning window function [22,23] and a global electron reduction factor of $S_0^2$=0.8. The disorder and high-frequency thermal vibration of the atoms was accounted for by a Debye-Waller term that is proportional to the difference of the mean square atomic displacements.

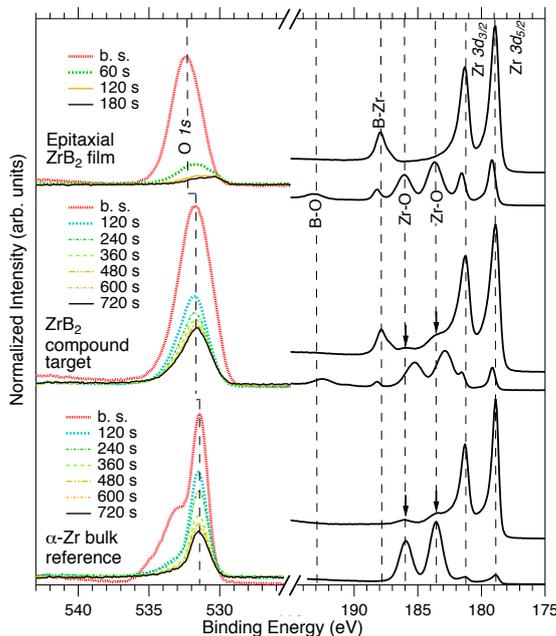

**Figure 1:** (Color online) Left panels: O *1s* XPS spectra before sputtering (b.s.) and after sputtering of the ZrB$_2$ epitaxial film (top), the ZrB$_2$ compound target (middle), and the α-Zr bulk reference (bottom). Right panels: Zr *3d* and B *1s* XPS spectra from the ZrB$_2$ epitaxial film (top), the ZrB$_2$ compound target (middle), and the α-Zr bulk reference (bottom) with spectra prior to sputter-cleaning at the bottom and after completed sputter-cleaning at the top. The vertical dashed lines are guides for the eye.

### 3. Results and Discussion

Prior to analysis by EXAFS and XANES, we determined the chemical bonding structure, composition, and structural properties of the three samples investigated. From the peaks of high intensities in the high-resolution O *1s* spectra (in the left side of Figure 1) and peaks attributed to Zr-O bonding in the high-resolution Zr *3d* XPS spectra (in the right side of Figure 1 and spectra at the bottom), it is evident that all samples exhibit a surface oxide prior to sputter-cleaning. In addition, the Zr *3d* spectra show the evidence for B-O bonds present in the epitaxial ZrB$_2$ film and in the ZrB$_2$ compound target seen from broad peaks of low intensities at positioned at binding energies at around 193 eV as visible in each respective spectrum. This binding energy is in agreement with the value 193.3 eV that was determined for B$_2$O$_3$ in [25] as well as our observation for ZrB$_2$ films deposited on 4H-SiC(0001) [3].

Furthermore, the peak at 193 eV is shifted by 5 eV eV towards higher binding energy with respect to the Zr-B signal present at ~188 eV, which is consistent with the higher electronegativity of O compared to B.

Sputter-cleaning reduces the intensity of the O *1s* peaks, particularly for the epitaxial $ZrB_2$ film. However, O *1s* peaks are still visible for all samples even after the last sputter-cleaning step has been completed, with the highest intensity in the α-Zr bulk reference followed by the $ZrB_2$ compound target, and the epitaxial $ZrB_2$ film.

The Zr *3d* spectra possess Zr-O peaks in both reference materials, while no such peaks are visible in the epitaxial $ZrB_2$ film, see spectra at the top in the right side of Figure 1. As listed in Table I, the binding energies (BE) of the Zr $3d_{5/2}$, $3d_{3/2}$ and B *1s* core-level peaks in the epitaxial $ZrB_2$ film are: 178.9 eV, 181.3 eV and 187.9 eV, respectively. These BEs are in agreement with those determined for epitaxial $ZrB_2$ films deposited on 4H-SiC(0001) in [3] with values of 179.0 eV, 181.3 eV and 188.0 eV, respectively. Furthermore, the binding energies are consistent with reported values for single-crystal bulk $ZrB_2$: 178.9 eV, 181.3 eV and 187.9 eV by Aizawa *et al.* [26] and 179.0 eV, 181.4 eV and 188.0 eV by Singh *et al.* [27]. For the $ZrB_2$ compound target the Zr *3d* and B *1s* main peaks are located at 178.9 eV, 181.3 eV and 187.9 eV, respectively, *i.e.* at the same BE as for the epitaxial film, and in agreement with the literature. The bulk α -Zr reference sample has the Zr $3d_{5/2}$ and Zr $3d_{3/2}$ peaks at 178.9 eV and 181.3 eV. These values are close to those determined for sputtered α-Zr films with 178.8 eV and 181.2 eV [28] and the study by Nyholm and Mårtensson on bulk Zr with 178.79 eV and 181.21 eV [29]. From the binding energies of the Zr $3d_{5/2}$ and Zr $3d_{3/2}$ peaks in the bulk α-Zr reference sample and the two $ZrB_2$ samples, we note a small to negligible shift to higher binding energy of 0-0.1 eV for Zr-B bonds compared to Zr-Zr bond. Thus, XPS suggests a limited change transfer from Zr to B in $ZrB_2$. This observation is supported by results from XPS valence band measurements, suggesting the bonding in $ZrB_2$ to be a combination of the graphitic bonding model in the planar B network and the hcp-metal bonding model for Zr [30].

Quantitative analysis following completed sputter cleaning as presented in Table II shows an oxygen content in the α-Zr bulk reference of ~22 at.% with a slightly lower content of ~19 at.% in the $ZrB_2$ compound target, and with the lowest oxygen content in the epitaxial $ZrB_2$ film of ~5 at.%. The C concentration follows a similar trend as the O content, and varies from more than 13 at.% in the α-Zr bulk reference, to ~8 at.% in the $ZrB_2$ compound target, and ~2 at.% in the epitaxial $ZrB_2$ film. It is clear from Table I that the investigated samples contain O and C contaminants with higher amounts in the reference target material compared to the epitaxial $ZrB_2$ film. This is caused by the polycrystalline character of the target which facilitates oxygen diffusion along grain boundaries upon prolonged air exposure. Such phenomena are very limited in the case of epitaxial film resulting in much lower O content, which is predominantly caused by the artefacts during Ar+ sputter-cleaning of surface native oxide.

Preferential sputtering and forward implantation are just two examples of phenomena that cause an enrichment of lighter elements such as O and C. In addition, re-deposition of sputtered contaminants on the $Ar^+$-etched surfaces during the time necessary to acquire the XPS spectra prevents complete surface cleaning [17]. Higher C and O content on the surfaces of Zr and $ZrB_2$ targets likely results from the high surface roughness which may prevent proper cleaning with the $Ar^+$ ion beam. One evidence supporting this interpretation is observed in the Zr *3d* spectra, that contain Zr-O peaks even after the last sputter-cleaning step. This is in contrast to the Zr *3d* spectrum of the epitaxial film, where the Zr-O peaks completely disappear after surface cleaning. The remaining O *1s* intensity in the latter case is due to re-deposited oxygen-containing species, but there is no evidence for Zr-oxide. For $ZrB_2$ films, time-of-flight energy elastic recoil detection analysis yields much lower O content than XPS, with ~1 at.% and even lower C content of ~0.4 at.% in the bulk of the investigated films [31]. It is therefore likely that XPS given the artefacts described above exaggerates the O and C content in all investigated samples, but the fact remains that the lowest content of contaminants is encountered in the epitaxial $ZrB_2$ and that this will affect the electronic structure and bonding distances determined by XANES and EXAFS.

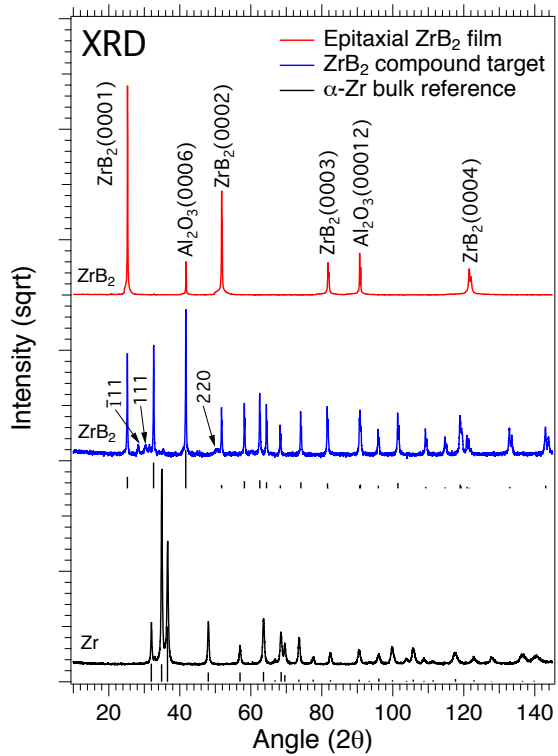

**Figure 2:** X-ray diffraction from the $ZrB_2$ epitaxial film (top), the $ZrB_2$ compound target (middle), and the α-Zr bulk reference. The bars below the $ZrB_2$ compound target and bulk reference α-Zr diffractograms represent peak positions and intensities from refs. [31] and [33], respectively. Note the square root intensity scale applied in the diffractograms.

Figure 2 shows XRD patterns from the epitaxial $ZrB_2$ film (diffractogram at the top), the $ZrB_2$ compound target (diffractogram in the middle) as well as for the α-Zr bulk reference (diffractogram at the bottom). As can be seen, the film demonstrates 000ℓ peaks (ℓ=1, 2, 3, and 4) of high intensities from the $ZrB_2$ phase [32] and where the 0001 and 0002 peaks exhibit higher intensities than the 0006 and 00012 peaks from the $Al_2O_3$(0001) substrate and with no visible peaks from other phases such as an oxide. There is a weak $ZrB_2$ $10\bar{1}0$ peak (not visible given the applied square root scale) showing a minority orientation in the film [4]. Furthermore from XRD pole figure measurements, the epitaxial relationships between the film and the substrate were determined to be in the out-of-plane direction $ZrB_2$(0001) ∥ $Al_2O_3$(0001) and with two in-plane relationships $ZrB_2[10\bar{1}0]\|Al_2O_3[10\bar{1}0]$ and $ZrB_2[11\bar{2}0]\|Al_2O_3[10\bar{1}0]$ [4]. The diffraction pattern from the $ZrB_2$ compound target displays all prominent $ZrB_2$ peaks listed in the JCPDS card for the investigated 2θ region and where the intensity distribution among the $ZrB_2$ peaks support a randomly oriented target material [32]. In addition, the diffractogram shows three weak, but visible peaks at the 2θ angles 28.42, 31.53, and 50.24 degrees. These are the $\bar{1}11$, 111, and 220 peaks in monoclinic $ZrO_2$ (m-$ZrO_2$) [33]. The formation of an oxide is supported from XPS revealing an O content of close to 19% in the $ZrB_2$ compound target after prolonged sputter-cleaning. This reflects the high affinity of Zr to O resulting in the oxidation of $ZrB_2$. The oxidation behavior of $ZrB_2$ has been the subject for extensive research seen from the materials properties as an ultra-high temperature ceramic, see e.g. [1] and references therein. A further fact is that the applied $ZrB_2$ target is a from $ZrB_2$ powder sintered body that is likely to contain pores in which contaminants as O and C can be dissolved during manufacturing and later to be released during thin film synthesis, see [1] and discussion on densification and sintering techniques for $ZrB_2$. From this, we note that a material containing minority phases as well as high amounts of O and C will demonstrate different properties compared to a phase-pure material with a low level of contaminants when investigated by XANES and EXAFS and in our specific case it may affect the macroscopic properties of films deposited from such a material.

The diffractogram from the α-Zr bulk reference at the bottom of Figure 2 exhibits peaks from hexagonal closest packed α-Zr [34]. The unit cell of α-Zr is of Mg type structure (Strukturbericht notation A3) and with atoms in (0,0,0) and (⅓, ⅔, ½). A closer inspection reveals that the intensity distribution for the peaks in the diffractogram differs from that of a randomly oriented material [34]. For our reference, the 0002-peak that is positioned at a 2θ angle of 34.84 degrees displays the highest intensity value, which shows that the metal is 0001-oriented and not randomly oriented. Differently, from the $ZrB_2$ compound target there are no crystalline minority phases in the reference sample despite an even higher O content.

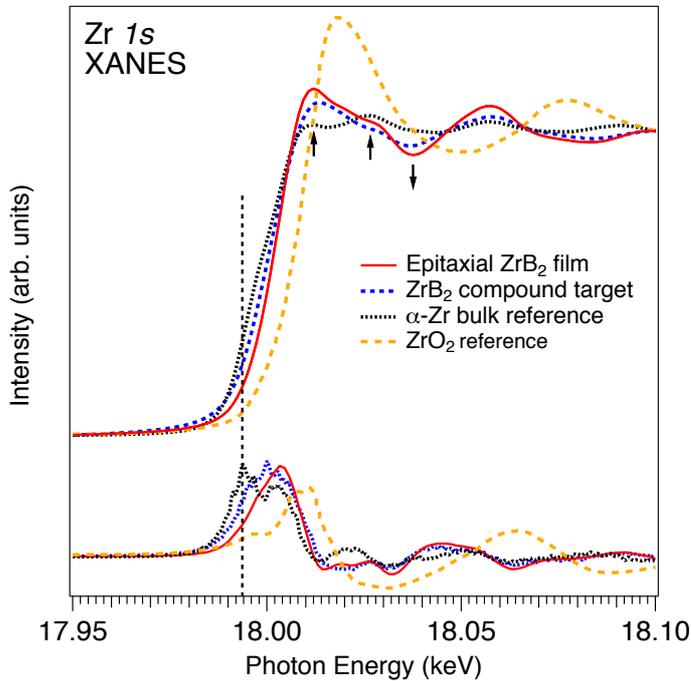

**Figure 3:** (Color online) Zr *1s* XANES spectra of the epitaxial $ZrB_2$ film, the $ZrB_2$ compound target and α-Zr bulk reference as well as $ZrO_2$ spectrum from A. Nozaki *et al.* [40]. The first derivative is shown at the bottom.

From the peak positions in the diffractograms in Figure 2, we determined the lattice parameters in the $ZrB_2$ compound target and the α-Zr bulk reference to *a*=3.167 Å and *c*=3.531 Å and *a*=3.233 Å and *c*=5.149 Å, respectively (see Table III). The lattice parameters for the epitaxial $ZrB_2$ film were determined from reciprocal space maps (RSM) to *a*=3.169 Å and *c*=3.528 Å, [4] and Table III. The measured values are close to the literature values with *a*=3.1687 Å and *c*=3.5300 Å for $ZrB_2$ [32] and *a*=3.232 Å and *c*=5.147 Å for α-Zr [34] that are also listed in Table III.

Figure 3 shows Zr *K* XANES recorded from the epitaxial $ZrB_2$ film, the $ZrB_2$ compound target, and the α-Zr bulk reference. The bottom part of Fig. 3 shows the first derivative of the absorption spectra presented above. The energy positions and the shapes of the main absorption peaks and the pre-edge XANES shoulders depend on the chemical state of the absorbing atom [35]. For the two $ZrB_2$ samples, the position of the absorption edge-step has moved to higher energy due to charge-transfer from the Zr to the B atoms. The high energy shift determined from the first derivative of the absorption edge (marked by the arrows) in comparison to pure α-Zr 17.994 keV (ref: 17.99273 keV) is 4 eV for the film and 6 eV for bulk. The energy shift is most pronounced for the bulk $ZrB_2$ target reference, which is related to a higher oxygen content as supported by Stewart *et al.* [16] for annealed e-beam co-evaporated thin films. We further note that an even larger shift of 10 eV was observed by Bösenberg *et al.* in their investigated $ZrB_2$ powder [15].

The main absorption peak of α-Zr is due to pure Zr *1s → 4p* dipole transitions forming a two-peak structure (indicated by arrows). The pre-edge shoulder (pre-peak) in the α-Zr spectrum is due to transitions into hybridized *p-d* states in elemental Zr, consistent with previous results [36,37]. The pre-peak is a signature of tetrahedral distortion of the coordination symmetry around the absorbing Zr atoms in the hexagonal α-Zr structure that allows *p-d* mixing into the Zr $1s \rightarrow 4p$ dipole transitions [38,39]. For the epitaxial $ZrB_2$ film and the $ZrB_2$ compound target, there is no hybridization between *p* and *d*-states and therefore the pre-peak is absent [40]. As observed, the amplitude of the oscillations starting at 18 keV are more pronounced in the $ZrB_2$ film than in both bulk materials. The generally sharper features in the epitaxial $ZrB_2$ film compared to both the $ZrB_2$ compound target and the bulk α-Zr reference is due to the high atomic ordering within the columns in the film, yielding well-defined directional bonds. Note that in XANES, local short-order unoccupied electronic structure is measured while XRD is a long-order probe. Therefore, it is important to use single-phase samples preferably with high ordering to determine the local chemical bonding structure.

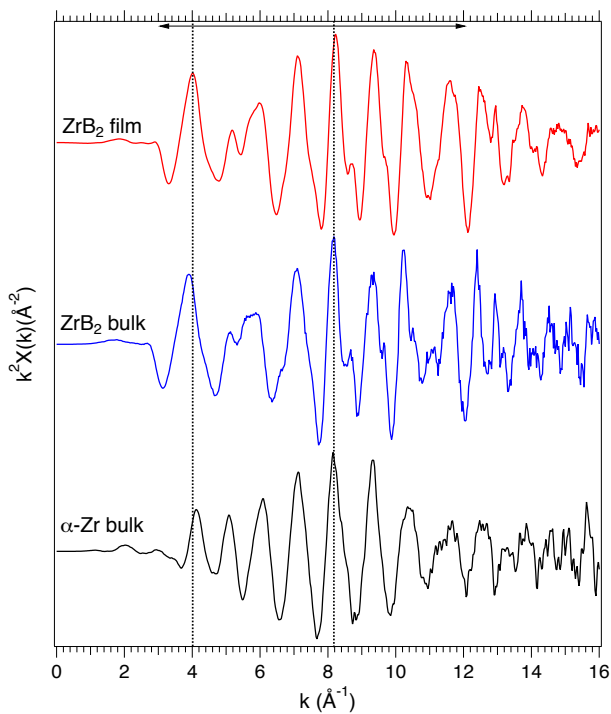

**Figure 4:** (Color online) EXAFS structure factor data S(Q) of the epitaxial $ZrB_2$ film at the top, the $ZrB_2$ compound target in the middle, and α-Zr bulk reference at the bottom. The horizontal arrow at the top shows the *k*-window for the most pronounced oscillations.

Comparing our recorded XANES spectra to those of Bösenberg *et al.* [15] and Stewart *et al.* [16], we find the same spectral features and peak shapes. Metallic Zr exhibit a main double-peak while $ZrB_2$ has a rather deep minimum as indicated by the arrows in Figure 4. The impurities in the bulk samples give rise to broader peaks in the spectra with lower intensities. Contrary to α-Zr, the main peak of the $ZrO_2$ reference spectrum from A. Nozaki *et al.* [41] has a broad and more intense single-peak shape that occur at higher energy and does not appear in the spectrum of the epitaxial $ZrB_2$ film. This shows that our film is of higher purity compared to $ZrB_2$ materials investigated in previous studies [15,16].

Figure 4 displays EXAFS structure factor oscillations of the epitaxial $ZrB_2$ film, the $ZrB_2$ compound target from which the film was synthesized, and the α-Zr bulk reference, obtained from raw data that has not been phase shifted. The structure factors $\chi$ are plotted as a function of the wave vector *k*, that were $k^2$-weighted to highlight the higher *k*-region, where $k = \hbar^{-1}\sqrt{2m \cdot (E - E_0)}$ is the wave vector of the excited electron in the X-ray absorption process. The oscillations were obtained after absorption edge determination, $E_o$ energy calibration by the first derivative, background subtraction and normalization to a spline function. The frequency of the oscillations and intensity of the EXAFS signal are directly related to the bond length (*R*) and the number of nearest neighbors (*N*), respectively. A higher frequency of the oscillations implies larger *R* while a larger amplitude implies increased *N*. For the investigated samples, the main sharp oscillations occur in the 3-12 Å$^{-1}$ *k*-space region, where the applied *k*-window is indicated by the horizontal arrow at the top of Fig. 4. Note the similarities and differences in the positions and the envelopes of the oscillations between the epitaxial $ZrB_2$ film and both bulk materials. The difference is most pronounced between the film and the α-Zr reference sample. In Figure 4, the main oscillation at 8.2 Å$^{-1}$ is due to Zr-Zr in-plane scattering and corresponds to the distance for the *a*-axis in the hexagonal crystal structure. It is similar in shape for all spectra, but where the peak becomes slightly shifted to higher *k*-values, in particular, for the thin film sample. This indicates shorter Zr-Zr bonds, which is to be expected when comparing the *a*-axes in respective crystal structure with ZrB$_2$: *a*=3.1687 Å [32] and Zr: *a*=3.232 Å [34] as well as Table III. The feature at 5.6-6.0 Å$^{-1}$ and the shoulder at 8.6 Å$^{-1}$ for the B-containing samples are associated with superimposed oscillations from Zr-B scattering. In the *k*-space region, the peak with the lowest *k*-value occurs at ~3.8 Å$^{-1}$ for the ZrB$_2$ compound target while it is shifted to ~4.0 Å$^{-1}$ for the epitaxial ZrB$_2$ film. This is an indication that the Zr-B bond is shorter in the epitaxial film than in the compound target. As observed in the α-Zr bulk reference sample, there is a superimposed shoulder originating from oscillations with low intensity originating from Zr-Zr scattering in this *k*-region. To analyze the detailed local structure and bond distances in the films, modeling of the raw EXAFS data was performed as shown below.

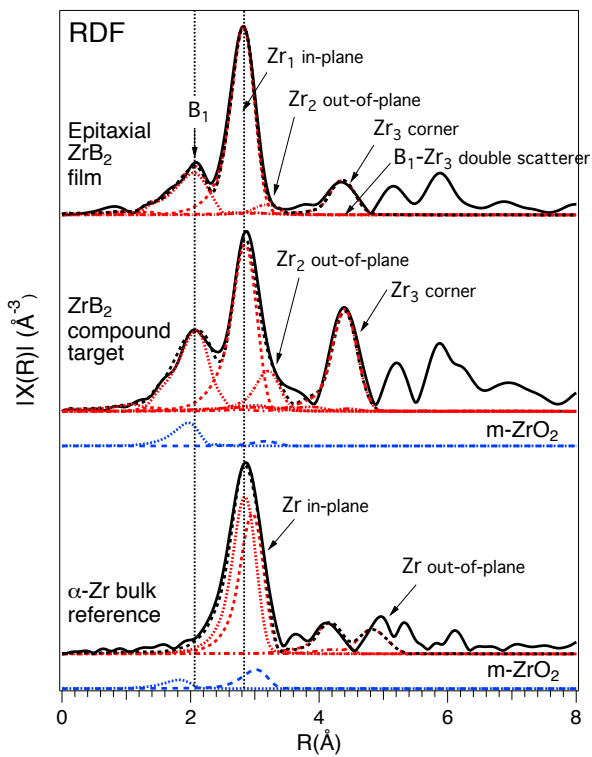

**Figure 5**: Radial distribution functions of the epitaxial ZrB$_2$ film at the top, ZrB$_2$ compound target in the middle, and α-Zr bulk reference at the bottom. Note that the peaks correspond to raw data from Fig. 5 that have not been phase shifted (~0.4 Å).

Figure 5 shows the magnitude of the radial distribution functions (RDFs) obtained from the raw data in Fig. 4, using Fourier transformation of the $k^2$-weighted $\chi(k)$ by the standard EXAFS procedure [42]. The quantitative analysis was made as described in section 2.2. Table IV shows the results of the EXAFS fitting using the FEFF scattering paths of ZrB$_2$, α-Zr and ZrO$_2$ as model systems. Note that the EXAFS data of the bulk ZrB$_2$ and α-Zr samples cannot be fitted without considering superimposed ZrO$_2$ in the modelling as in the case of Stewart *et al.* [16] where tetragonal ZrO$_2$ was applied. In our study, the XRD pattern recorded from the ZrB$_2$ compound target in Fig. 2, shows m-ZrO$_2$. Thus, we use m-ZrO$_2$ in our modelling although EXAFS cannot distinguish between different ZrO$_2$ polytypes. The bond lengths (R) are compared to those obtained from the lattice parameters determined from XRD in Table III. Firstly, we compare the atomic distances of the α-Zr bulk reference with the B-containing samples.

For α-Zr [34], the main peak is dominated by the Zr-Zr paths at 3.160 Å (half diagonal, same notations as in Fig. 5) and 3.254 Å (cell edge), both with six nearest neighbors in the first coordination shell, which is close to our values obtained from θ/2θ XRD of the α-Zr bulk reference with 3.180 Å and 3.233 Å, respectively [37] [43] and Table III. Our XRD data is in excellent agreement with the bond lengths (3.179 Å and 3.232 Å) calculated from the JCPDS card [34]. Here, the bond lengths from XRD and literature values are determined for specific lattice parameters in Table III. The Zr-Zr scattering path of the *c*-axis is observed as a weak feature at 5.147 Å to compare with 5.149 Å from our XRD measurements.

Comparing the two $ZrB_2$ samples, we find the cell-edge Zr-Zr distances of the in- and out-of-plane contributions (peaks $Zr_1$ and $Zr_2$ in Fig. 5) at lower distances in the epitaxial $ZrB_2$ film (3.165 Å and 3.520 Å) than the Zr-Zr bond distances in the $ZrB_2$ target (3.175, 3.540 Å). Note that we compare the cell edges (*a* and *c*-axis) since the half-diagonal is replaced by Zr-B bonds in the $ZrB_2$ structure. The calculated bond length from literature lattice parameters is 3.169 Å [32] that is close to the epitaxial film, as listed in Table IV. The RSM recorded from the epitaxial $ZrB_2$ film shows 3.169 Å and θ/2θ from the target yields 3.167 Å. The $Zr_3$ corner scatterer has a slightly shorter distance but significantly lower intensity in the epitaxial film than the $ZrB_2$ compound target due to the films orientation along the *c*-axis.

Secondly, the Zr-B distance (peak $B_1$ in Fig. 5), is somewhat shorter (2.539 Å) in the epitaxial $ZrB_2$ thin film than in the bulk sample (2.599 Å). The Zr-B bond distance is larger in bulk than in the thin film sample due to additional superimposed Zr-O bonds, which is consistent with the observations in XANES, XPS, and XRD. The Zr-B value (2.539 Å) is in better agreement with the calculated bond distance from the lattice parameters of the bulk literature value (Zr-B=2.542 Å) [32] than the $ZrB_2$ compound target listed in Table III. Furthermore, the Zr-B bond length in the $ZrB_2$ film is shorter than that obtained by Stewart *et al.* [16], with (2.546 Å) and Chu *et al.* [13] [14] (2.55 Å) as well as the calculated bond length of stoichiometric $ZrB_2$ by Lee *et al.* [44] (2.81 Å). A longer Zr-B bond is likely due to additional impurities and non-directional bonds.

From our results, it is evident that the epitaxial $ZrB_2$ film exhibits superior electronic structure properties in terms of chemical bonding alignment, crystal quality and level of contaminants when compared to the $ZrB_2$ compound target from which it was synthesized. The combined XANES and EXAFS studies reveal several interesting observations. For the XANES spectra, the inclusion of boron has a large effect on the unoccupied electronic structure and spectral shape of the main absorption peak. However, when oxygen is present, there is a broadening and smearing of the spectral features as adventitious oxides often are X-ray amorphous. Moreover, the EXAFS analysis show that the bond distances are also affected by oxygen by expanding the Zr-Zr bonds. The Zr-B bonding is longer in the bulk material than in the thin film, which is attributed to less ideal bond lengths where the contamination affects the chemical surrounding. Furthermore, due to the in-plane polarization of the X-rays, the epitaxial $ZrB_2$ film has a significantly higher coordination number contribution in the basal-plane than along the *c*-axis in the EXAFS spectra. The strong polarization dependence is a consequence of the 0001-orientation of the film to the $Al_2O_3$(0001) substrate [4]. It can be anticipated that future single-crystal $ZrB_2$ films will display even stronger polarization dependence. As a comparison, for the weakly 0001-oriented α-Zr bulk reference material, the polarization dependence appears to be more isotropic as the peak originating from the *c*-axis is very weak in EXAFS.

Another observation is the fact that EXAFS shows a larger difference in atomic distances than XRD between the film and bulk materials. The discrepancy between the EXAFS and XRD measurements are attributed to the characteristic of element-specific short-range order probe for the intrinsic coordination structure (bond distance, coordination number and disorder) in EXAFS while, XRD probes long range order, where the lattice constants define the dimensions of the unit cells. This put demands on the samples as our study shows that materials must have uniform structure and low level of contaminants that are important when investigating the electronic structure and bonding distances with XANES and EXAFS. Epitaxial films are therefore promising for property determination of borides with C32 structure and future exploration also of other material classes.

## Conclusions

From a combination of analysis with XPS, XRD, XANES, and EXAFS, we investigate the local chemical bonding structure and structural properties with atomic distances in an epitaxial $ZrB_2$ film and compare these properties to those of the $ZrB_2$ compound target from which the film was synthesized as well as a bulk $\alpha$-Zr reference. XPS shows that the film surface region contains the lowest amount of contaminants O and C with ~5 at.% and ~2 at.%, respectively. XRD reveals m-$ZrO_2$ as an minority phase in the $ZrB_2$ compound target, while no crystalline minority phases are detected in the bulk $\alpha$-Zr reference despite ~22 at.% O and ~13 at.% C as determined by XPS. The XANES data from the $ZrB_2$ film exhibits stronger oscillations and no oxygen contribution, indicating high sample quality. From EXAFS, the shortest Zr-Zr bond distances were found in the $ZrB_2$ thin film (3.165) while somewhat larger distances were found in bulk $ZrB_2$ (3.175 Å) and $\alpha$-Zr (3.254 Å). The Zr-B distance in $ZrB_2$ is also shorter in the thin film, 2.539 Å and 2.599 Å, respectively and the EXAFS data of the bulk $ZrB_2$ sample cannot be fitted without taking into account superimposed $ZrO_2$ in the modelling. These bond differences are thus due to oxygen in the compound bulk target. Therefore, epitaxial films serve as the better reference material.

## 7. Acknowledgements

We would like to thank the staff at MAX-IV Laboratory for experimental support. MM acknowledges financial support from the Swedish Energy Research (no. 43606-1), the Swedish Foundation for Strategic Research (SSF) (no. RMA11-0029) through the synergy grant FUNCASE and the Carl Trygger Foundation (CTS16:303, CTS14:310). LT acknowledges the Swedish Research Council (VR) through the contract 621-2010 3921. GG thanks the Knut and Alice Wallenberg Foundation Scholar Grant KAW2016.0358, the VINN Excellence Center Functional Nanoscale Materials (FunMat-2) Grant 2016-05156, and the Åforsk Foundation Grant 16-359. LH and HH acknowledge financial support from the Swedish Government Strategic Research Area in Materials Science on Functional Materials at Linköping University (Faculty Grant SFO-Mat-LiU No. 2009-00971).

**TABLE I:** Binding energies of the *3d$_{5/2}$*, *3d$_{3/2}$* and B *1s* core-level peaks. *This study.

| Sample | 3d$_{5/2}$ | 3d$_{3/2}$ | B *1s* |
|---|---|---|---|
| Epitaxial ZrB$_2$ film* | 178.9 | 181.3 | 187.9 |
| Epitaxial ZrB$_2$ film [3] | 179.0 | 181.3 | 188.0 |
| Single crystal bulk ZrB$_2$ [26] | 178.9 | 181.3 | 187.9 |
| Bulk ZrB$_2$ by Singh [27] | 179.0 | 181.4 | 188.0 |
| ZrB$_2$ compound target* | 178.9 | 181.3 | 187.9 |
| α-Zr bulk reference* | 178.9 | 181.3 | - |
| α-Zr films [28] | 178.8 | 181.2 | - |
| Nyholm and Mårtensson [29] | 178.79 | 181.21 | - |

TABLE II: Quantitative analysis by XPS after sputter-cleaning.

| Sample | B (at.%) | Zr (at%) | O (at.%) | C (at.%) |
|---|---|---|---|---|
| Epitaxial ZrB$_2$ film | 61.2±3.0 | 32.1±1.6 | 4.7±0.2 | 2.0±0.1 |
| ZrB$_2$ compound target | 41.3±2.0 | 32.2±1.6 | 18.6±1.0 | 7.9±0.4 |
| α-Zr bulk reference | - | 64.6±3.2 | 22.1±1.0 | 13.4±0.7 |

**TABLE III:** Lattice parameters of the investigated samples determined from XRD and from the JCPDS cards for ZrB$_2$ [31] and for α-Zr [33].

| Sample | *a* (Å) | *c* (Å) |
|---|---|---|
| ZrB$_2$ [31] | 3.169 | 3.530 |
| Epitaxial ZrB$_2$ film [12] | 3.169 | 3.528 |
| ZrB$_2$ compound target | 3.167 | 3.531 |
| α-Zr [33] | 3.232 | 5.147 |
| α-Zr bulk reference | 3.233 | 5.149 |

**TABLE IV:** Structural parameters for the epitaxial ZrB$_2$ film and ZrB$_2$ compound target in comparison to pure α-Zr bulk reference obtained from fitting of calculated radial distribution functions in the first coordination shell. *N* is the coordination number, *R* is the bond length (in Å) for the Zr-B and Zr-Zr, scattering paths respectively, σ is the corresponding Debye-Waller factor, representing the amount of atomic displacement and disorder, reduced $\chi_r^2$ as the squared area of the residual. Bond lengths from lattice parameters in XRD are given in parenthesis.

| System | Shell | R(Å) | N | σ(Å$^2$) | ΔE$_0$(eV) | Statistics |
|---|---|---|---|---|---|---|
| ZrB$_2$ film | Zr-B (N=12) | 2.539±0.005 (2.542) | 6.204±0.01 | 0.0020±0.001 | 1.08±0.01 | $\chi_{16}^{0.95}$ = 25.73 N=28, P=12 ν=N-P=16 |
|  | Zr-Zr (N=6) *a* | 3.165±0.005 (3.169) | 6.000±0.01 | 0.0021±0.001 | 0.58±0.01 |  |
|  | Zr-Zr (N=2) *c* | 3.520±0.005 (3.528) | 0.500±0.01 | 0.0022±0.001 | 1.08±0.01 |  |
|  | Zr-Zr (N=12) | 4.743±0.005 (4.743) | 3.828±0.01 | 0.0026±0.001 | 1.08±0.01 |  |
|  | Zr-B-Zr (N=48) | 4.949±0.005 (5.083) | 48.00±0.01 | 0.0049±0.001 | 1.08±0.01 |  |
| ZrB$_2$ target | Zr-B (N=12) | 2.599±0.005 (2.542) | 10.00±0.01 | 0.0045±0.001 | 0.10±0.01 | $\chi_{24}^{0.95}$ = 35.8 N=28, P=4 ν=N-P=24 |
|  | Zr-Zr (N=6) *a* | 3.175±0.005 (3.167) | 6.000±0.01 | 0.0046±0.001 | 0.11±0.01 |  |
|  | Zr-Zr (N=2) *c* | 3.540±0.005 (3.531) | 2.000±0.01 | 0.0048±0.001 | 0.10±0.01 |  |
|  | Zr-Zr (N=12) | 4.757±0.005 (4.743) | 12.00±0.01 | 0.0049±0.001 | 0.11±0.01 |  |
|  | Zr-B-Zr (N=48) | 4.800±0.005 (5.084) | 47.00±0.01 | 0.0049±0.001 | 0.11±0.01 |  |
|  | Zr-O (N=1) | 2.387±0.005 | 0.817±0.01 | 0.0020±0.001 | 0.11±0.01 |  |
|  | Zr-Zr (N=2) | 3.500±0.005 | 0.200±0.01 | 0.0047±0.001 | 0.11±0.01 |  |
| α-Zr | Zr-Zr (N=6) | 3.160±0.005 (3.180) | 6.00 | 0.0035±0.001 | 0.67±0.01 | $\chi_{8}^{0.95}$ = 14.06 N=12, P=4 ν=N-P=8 |
|  | Zr-Zr (N=6) *a* | 3.254±0.005 (3.233) | 6.00 | 0.0040±0.001 | 0.67±0.01 |  |
|  | Zr-Zr (N=6) | 4.496±0.005 | 6.00 | 0.0089±0.001 | 0.67±0.01 |  |
|  | Zr-Zr (N=6) *c* | 5.147±0.005 (5.149) | 2.00 | 0.0091±0.001 | 0.67±0.01 |  |
|  | Zr-O (N=1) | 2.260±0.005 | 0.40 | 0.0034±0.001 | 0.67±0.01 |  |
|  | Zr-Zr (N=2) | 3.300±0.005 | 0.80 | 0.0041±0.001 | 0.67±0.01 |  |